\documentstyle[PASJadd,psfig]{PASJ95}
\begin{document}

\title{ASCA Observations of the Temperature and Metal Distribution in
the Perseus Cluster of Galaxies}

\author{ H.~{\sc Ezawa},$^{1}$ Y.~{\sc Fukazawa},$^{2}$ M.~{\sc
Hirayama},$^3$ H.~{\sc Honda}, $^4$ T.~{\sc Kamae},$^{2}$ K.~{\sc
Kikuchi},$^{5}$ \\ T.~{\sc Ohashi},$^6$ R.~{\sc Shibata},$^7$ and
N.~Y.~{\sc Yamasaki}$^6$ \\ [12pt]
$^1${\it Nobeyama Radio Observatory, National Astronomical
Observatory, Minamimaki, Minamisaku, Nagano 384-1305, Japan}\\
 {\it E-mail(HE): ezawa@nro.nao.ac.jp}\\
$^2${\it Department of Physics, Hiroshima University, 1-3-1, Kagamiyama, Higash
ihiroshima, Hiroshima 739-8526, Japan} \\
$^3${\it Santa Cruz Institute for Particle Physics, University of
California, Santa Cruz, CA 95064, U. S. A.} \\
$^4${\it Department of Information Science, Shonan Institute of Technology,
Fujisawa, Kanagawa 251-8511, Japan}\\
$^5${\it Tsukuba Space Center, National Space
Development Agency (NASDA), Ibaraki 305-8505, Japan}\\ 
$^6${\it Department of Physics, Tokyo Metropolitan University, 1-1
Minami-Ohsawa, Hachioji, Tokyo 192-0397, Japan}\\
$^7${\it Institute of Space and Astronautical Science, 3-1-1 Yoshinodai, Sagami
hara, Kanagawa 229-8510, Japan}
} 

\abst{
Large-scale distributions of hot-gas temperature and Fe abundance in
the Perseus cluster have been studied with multi-pointing observations
by the GIS instrument on ASCA\@.  Within a radius of $20'$ from the
cluster center, the energy spectra requires two temperature
components, in which the cool component indicates $kT \sim 2 $ keV and
the hot-component temperature shows a significant decline from about 8
keV to 6 keV toward the center.  In the outer region of the cluster,
the temperature shows a fluctuation with an amplitude of about 2 keV
and suggest that a western region at $\sim 16'$ from the cluster center
is relatively hotter.  As for the Fe abundance, a significant decline
with radius is detected from $0.44$ solar at the center to $\sim 0.1$
solar at a $50'$ offset region. If observed Fe-K line intensity within
$4'$ from the center is suppressed by a factor of 2 due to the
resonance scattering effect, the corrected Fe mass density follows the
galaxy distribution. Finally, our results do not support the
large-scale velocity gradients previously reported from the same
GIS data.
}
\kword{galaxies:abundance --- galaxies:clusters:
individual (Perseus) --- intergalactic medium --- X-rays:galaxies}

\maketitle \thispagestyle{headings}

\section{Introduction}

The distributions of temperature and metal abundance in the
intracluster medium (ICM, hereafter) effectively constrain the
evolutionary history of clusters of galaxies. Significant temperature
structures have been detected with the ASCA for a number of clusters, some
of which show regular X-ray morphologies (see e.g.\ Markevitch et al.\
1998, Watanabe et al.\ 1999, Churazov et al.\ 1999, Furusho et al. \ 2000). 
This indicates
that the temperature features are the most effective tool in
detecting the past occurrence of subcluster mergers. Metal
distribution, on the other hand, tells us the injection mechanism from
galaxies and how much the ICM has been mixed. ASCA brings us the first
capability to map out the ICM with suitable energy range (0.7--10 keV)
and with enough energy resolution (Tanaka et al.\ 1994). Most of the
clusters for which temperature and metal distributions were reported
have been  covered in a single GIS field, therefore the objects had to be
somewhat distant and internal structures were not well resolved. For
very near-by clusters, mapping observations are inevitably needed and
the process of data analysis becomes complicated because of the
response function of the X-ray telescope (XRT)\@. The results reported
in this paper are based on multi-pointing observations of the Perseus 
cluster with ASCA.

Perseus cluster is the brightest cluster of galaxies with a large
angular extent ($\gtsim 2^\circ$ in diameter) and also the strongest
iron-line emitter in the X-ray sky.  Observational study of the
temperature and metal distribution has a long history. Spacelab 2
(Ponman, et al.\ 1990) and Spartan 1 (Kowalski et al.\ 1993) showed
that the ICM temperature was mostly uniform but a significant
concentration of Fe around 1 solar is present toward the cluster
center. These instruments relied on an arithmetical reconstruction of
the image from the scanning or coded-mask data. BBXRT observation
showed almost constant Fe abundance at $\sim 0.5$ solar in the central
$10'$ region (Arnaud et al.\ 1992). The Beppo SAX data showed the abundance 
distribution within a radius of $20'$ (De Grandi, Molendi 2000) and 
indicated a  reduction of the Fe abundance within $4'$ from the cluster 
center due to a ``resonance scattering'' of Fe-K$\alpha$
line (Molendi et al.\ 1998). This interpretation was however questioned
by Dupke and Arnaud (1999). The surface brightness distribution to a
radius of $80'$ was studied by Schwarz et al.\ (1992) and Ettori et
al.\ (1998) based on the ROSAT PSPC data, and an excess emission over
a symmetric $\beta$-model profile is found at $20'-40'$ in the
east. This feature suggested a merging subcluster. An earlier ASCA
observation indicated a high temperature region with $kT \sim 10$ keV
at about $20'$ in the north (Arnaud et al.\ 1994).
The detailed X-ray morphology around the central galaxy NGC1275 
hosting the radio source 3C 84 was studied by ROSAT HRI 
(B\"ohringer et al. \ 1993; Churazov et al. \ 2000)
and most recently by the Chandra observatory (Fabian et al. \ 2000).
Dupke and Bregman (2001) recently reported the velocity gradient 
in the ICM by 1000 km s$^{-1}$ based on the same GIS data.

This paper reports ASCA results on the  large scale distributions of 
temperature and metal abundance 
based on multi-pointing observations. An $H_0 = 50$ km s$^{-1}$
Mpc$^{-1}$ and $q_0 = 0.5$ is employed, giving   31.5 kpc per  $1'$ at
the Perseus cluster, and a number fraction of ${\rm Fe}/{\rm H} = 4.68
\times 10^{-5}$ (Anders, Grevesse 1989) is used as 
the 1 solar Fe abundance.

\section{Observations and Analysis}
ASCA observations of the Perseus cluster have been performed with a
total of 13 pointings for an average exposure time of 15 ks
each. Table 1 shows the log of the ASCA observations. A region of
approximately $2^\circ \times 2^\circ$ was covered with the GIS (see
figure 1).  All the data were screened with the standard criterion,
such as cutoff rigidity greater than 8 GV and elevation from the earth
rim exceeding $5^\circ$, and flare-like events were excluded using 
H0 and H2 monitor counts (Ishisaki 1996). 
The long-term variation of the non X-ray background was also
considered in the subtraction process. Dead
time correction was applied for the central pointing data, which had
an intensity of about 10 c~s$^{-1}$ GIS$^{-1}$.  As for the X-ray
background, data from 20 blank sky regions after excluding point
sources were combined and used (Ikebe 1995).

A mosaic image of the Perseus cluster was constructed by projecting all
the observed data in the sky, as shown in figure 1. Significant X-ray
emission was detected even at the outermost boundary ($r \sim 60'$) of
the observed region. Two point sources are recognized in the
south-west direction of the cluster center. The one closer to the
cluster center is the radio galaxy IC 310 ($40'$ from the cluster
center), and the other source is 1RXS J031525.1+410620 which is not
optically identified yet.

In this paper, we only deal with the GIS data, since the solid angle
covered by the GIS in each pointing is larger than the SIS by a factor of
about 5 and the detector performance has been extremely stable since
the launch (Ohashi et al.\ 1996; Makishima et al.\ 1996;
Ishisaki et al. 1997). The data
from GIS2 and GIS3 were  combined and treated as a single pulse-height
spectrum.

As discussed in several papers (Takahashi et al.\ 1995; Honda et al.\
1996; Ikebe et al.\ 1997), the extended point-spread function and
stray light due to the XRT is a severe problem in the analysis of
cluster data. The spectral data in a certain detector region are
always contaminated by a contribution from near-by surrounding regions
due to the tail of the PSF\@. Contamination from a large offset angle
($\sim 1^\circ$) also exists, called as the stray light. 

For the conservative estimation of spectral parameters, we will
analyze the mapping data using the ``isothermal'' response
function. This method gives an approximate values of spectral
parameters by employing the instrument response function for a
constant temperature and metal abundance all over the cluster (see
Honda et al.\ 1996; Kikuchi et al.\ 1999). This assumption enables
us to estimate the spectral parameters for individual regions
separately. The only a priori information needed for the analysis is
the surface brightness distribution. As for the template image, we
employed  the ROSAT PSPC data of the Perseus cluster in the hard energy band
(0.5 -- 2 keV)\@. These data are described in Ettori et al.\
(1998). Since the image quality of the PSPC data is much better than
the ASCA one, the observed raw image is fine enough to be used as the
template.

\section{Results}
\subsection{The central region}
The energy spectrum for the central region of the cluster within $r = 4'$ was
examined first. The background was first  subtracted from the 
pulse-height spectra of GIS2 and
GIS3\@. We fit the combined GIS pulse-height spectra in the energy
range 0.7--9 keV with the MEKAL model, which reproduces the Fe-L line
feature better than other models. All the elements were assumed to
have the solar abundance ratio. 
We have confirmed
that the GIS response function needs a systematic error of 0.5\% to
obtain an acceptable fit to the Crab nebula spectrum, which is used as
the spectral standard. Also, our XRT response calculation based on the
Monte-Carlo simulation leaves an error of about 1\% in each energy
bin. 
As a sum of these errors we employed a systematic error of 1.2\%
in all pulse-height channels which was  added to the statistical error
because $1\sigma$ statistical errors are only 1--2\% of 
the measured intensity.
The detector gain was also adjusted to give a satisfactory fit for 
the Fe-K line of the Perseus cluster.
The best fit value of the gain parameter is 0.996$\pm$0.03 for the 
combination of the mapping data, and 
it is within  the accuracy of 1\%  derived by the GIS 
calibration (Makishima et al. \ 1996).

We first fitted the spectrum with a single-temperature MEKAL model, in
which the metal abundance and the interstellar absorption were varied
as free parameters. The model was found to be unacceptable with the
best-fit temperature $kT = 4.09$ keV and $\chi^2 = 304.5$ for 123
degrees of freedom (DOF) as shown in table 3. The residuals of the fit
are shown in the bottom panel of figure 2(a).  There is a clear
discrepancy between the data and the model around 1 keV, most probably
due to the presence of Fe-L emission lines, which  suggest that a
cool-temperature component with $kT \ltsim 2$ keV exists in the energy
spectrum.

We then tried 2-temperature models. In the 2-temperature fits, we can
in principle vary 8 parameters independently: which are intensity,
temperature, metal abundance, and absorption for each
component. However, since the GIS is rather insensitive to the
low-energy absorption below 0.7 keV, allowing all the parameters to vary
freely is not practical in constraining the spectral model. In order to
see how stable  the spectral parameters are for different combinations
of free and fixed parameters in the fit, we investigated the dependency 
of the fitting  models for
this central-region spectrum. In particular, we looked into the effect
of $N_{\rm H}$ and metal abundance by assuming them as either common
or independent parameters between the hot and cool components.

All the fitting results we have tested are shown in table 2.  The cool
component shows fairly large change of parameters among the fits. When
$N_{\rm H}$ of the cool component was forced to be the same as the
hot-component level, its metal abundance became lower than the
hot-component level and the temperature was constrained between 1.3 --
2.1 keV\@.  When the cool component was allowed to have its own free
$N_{\rm H}$ by fixing the hot-component absorption  at the Galactic
value ($1.4 \times 10^{21}$ cm$^{-2}$), the cool component indicated a
large abundance error (0.33--1.1 solar) and about 3 times higher
$N_{\rm H}$ than the Galactic level. The minimum $\chi^2$ value was
132.3 for 119 DOF, indicating that the fit was acceptable with the
90\% confidence. The reduction of the $\chi^2$ value by 15.7 from the
common $N_{\rm H}$ case (2nd row of table 2) was formally significant
at more than the 99\% confidence. We note that $N_{\rm H,cool}$ was
still high when we allowed a separate metal abundance for the cool
component ($Z_{\rm cool}$). It seems, therefore, likely that the cool
component has its own absorber, however we should be careful because
the GIS sensitivity is very low below 0.7 keV (Ohashi et al.\
1996). Fabian et al.\ (1994) also reported an excess absorption of
$1.1 \times 10^{21}$ cm$^{-2}$ based on the 2-temperature fit of the
SIS data. Recent Chandra color map at the cluster core also indicated
an excess absorption (Fabian et al. 2000). To compare with the results
for the outer regions, the model with fixed $N_{\rm H}$ at the
Galactic value and common metal abundance (the 1st row in table 2) was
fitted to the data in figure 2(a). The residuals of these
2-temperature fits are shown in the middle panel. The improvements in
the fit around $E\sim 1$ keV and 6.7 keV from the 1-temperature fit
are apparent.  This fit gives somewhat large $\chi^2$ value but
essentially the same hot-component parameters with the other fits.

We note that the temperature and metal abundance of the hot component
show rather small change for different fits. The temperature changes
by 10\% (5.23 to 5.73 keV), and the abundance by 7\% (0.43 to 0.46
solar), respectively. The 90\% confidence limits for the hot-component
parameters for the 6 fits in table 2 overlap, while the temperature of
the cool component indicates a significant change at the 90\%
confidence. Therefore, as far as the hot component properties are
concerned, we consider that the assumption of the common $N_{\rm H}$
and common abundance have little effect on the fitting results.  At the
same time, we expect that the true systematic error must be larger
than the formal error shown in table 2, because the 2-temperature
model itself should be regarded as a simplified approximation of a
multi-temperature mixture.  However, regarding the recent XMM-Newton finding
of the lack of low-temperature gas ($kT \leq 1.5$ keV) in A1795
(Tamura et al.\ 2001) and in A1835 (Peterson et al.\ 2001), the cool
component derived here may be related to such minimum temperatures
found in the cluster center.

The central galaxy NGC 1275 hosts a radio source 3C 84 which shows a
prominent radio lobe. To constrain the possible contribution from the
active nucleus, a power-law component was included in the model
spectrum. The photon index was fixed to $\Gamma = 2.1$ based on the
result by Ponman et al.\ (1990). Assuming the 2-temperature model for
the thermal emission (the 1st row of table 2), 
the 90\% upper limit for the power-law flux become
$7.1 \times 10^{-12}$ erg cm$^{-2}$ s$^{-1}$ in 0.7--2 keV ($2.7
\times 10^{-12}$ erg cm$^{-2}$ s$^{-1}$ in 2--10 keV). This
corresponds to $L_X < 1.5 \times 10^{43}$ erg s$^{-1}$ in 2--10 keV,
indicating the non-thermal flux to be  less than 1\% of the total cluster flux
and about $1/10$ of the level reported by Ponman et al.\ (1990). This
result is consistent with the previous Ginga value by Allen et al.\
(1992), who reported that the power-law component was less than 2\% of
the cluster flux.

\subsection{Radial variation}

The pulse-height data of the mapping observations were accumulated in
8 concentric annular regions in the radial range of $0' - 60'$ from
the cluster center, from which bright sources (IC310 and 1RXS
J031525.1+410620) were removed. Figure 3 shows examples of the
background subtracted spectra in 4 annular regions for the sum of the
GIS2 and GIS3 detectors. These spectra apparently indicate gradual
decline in the Fe-K line equivalent width as a function of the radius.

These annular pulse-height spectra were then fitted with the
2-temperature model.  As shown in table 3, the inclusion of the cool
component gave a reduction of the $\chi^2$ value by 154.6, 98.7, 40.8,
32.2, 12.0 for a change of 2 degrees of freedom for the radii $0'-4',
4'-8', 8'-12', 12'-20'$, and $20'-28'$, respectively, implying the
cool emission to be significant at more than the 99\% confidence
(see Lampton et al. 1976). 
In figure 2(b), the fitting result for the
annular region of $12'<r<20'$ is shown as an example,in which the residuals
are for 2-temperature (middle panel) and single-tempelarature (bottom-panel)
fits, respectively.
 In this series
of spectral fits, we assumed common $N_{\rm H}$ and metal abundance
for the two components based on the following reasons. The fits for
the central-region spectrum showed that the common abundance is a
reasonable assumption. Even though the free-$N_{\rm H}$ models for the
cool component indicated 3 times higher absorption, the outer regions
are likely to have only the Galactic $N_{\rm H}$\@. We also note that
the lower statistics of the data in the outer regions hamper accurate
determination of $N_{\rm H}$ and abundance for the cool component
separately. The resultant spectral parameters and their 90\%
confidence limits are listed in table 3 and their radial variations
are shown in figure 4 with filled circles. Results from single-temperature 
fits are also shows in table 3 and figure 4 for comparison.

One notable feature is that the hot spectral component shows a
significant decline of temperature in the inner regions ($r < 8'$ or
250 kpc). The central ``hot'' temperature, 5.73 keV (less than 6.23
keV at the 90\% confidence), is much lower than the outer values of
about 8 keV at a radius of $10' - 30'$. Since all the models fitted to
the central spectrum in the previous section indicate the ``hot''
temperature to be less than 6.2 keV, it seems certain that the gas is
cooling in the central region where the gas density is high.  We need 
to note that the
radial temperature profile in figure 4 is more extended than the true
feature because of the point spread function.  
Ikebe et al.\ (1999) carried out the
similar 2-temperature fit for the Centaurs cluster and found that the
temperature of the hot component was constant with radius.

The cool spectral component with $kT \sim 2$ keV shows a large spatial
extent and carries more than 20\% of the total
flux in the range $r = 0'-28'$ and then drops to less than 10\% in the
outer regions, where the significance of the cool emission also
becomes less than 90\%. The fraction of the cool-component in the
energy range 0.7 -- 2 keV was obtained for each annular region and
shown in table 3. By multiplying the cool component fraction to the 
PSPC flux in each annular region,
the surface brightness profile of this  component was calculated and 
shown in figure 5, which indicates that the emission is more extended
than the point-spread function. The luminosity of the cool component
is ($1.9\pm0.5)\times 10^{44}$ erg s$^{-1}$ in 0.7--2 keV and 
($3.7\pm0.9)\times 10^{44}$ erg s$^{-1}$ in 0.1--10 keV, respectively, 
which is comparable to the total luminosity of fairly rich clusters.

As for the metal abundance, a slow decline as a function of radius is
clearly seen in figure 4. The fit with the 2-temperature model
indicates that the metal abundance is 0.44 (0.41--0.47) solar within
$r = 4'$ and drops to 0.18 (0.10--0.27) solar at the outermost annular
region of $r = 44'-60'$, where $60'$ corresponds to 1.8 Mpc at the
source. Assumption of a constant abundance is rejected with $\chi^2
=358.3$ for 7 DOF for the best-fit value of 0.255 solar. The abundances
within $r=20'$ are consistent with the BeppoSAX results (De Grandi and
Molendi 2000). Since Fe-K emission line essentially determines the
abundance parameter, this feature does not change much ($< 10\%$) with
the choice of 1 or 2 temperature models. The actual abundance profile
must be steeper, but it is certainly broader than the point-spread
function. Also, we note that the central abundance increase
is not as strong as that in the Centaurs cluster (Fukazawa et
al.\ 1994; Ikebe et al.\ 1999). As seen in the pulse-height spectra in
figure 3, the relatively high temperature of the Perseus cluster makes
the quantitative determination of the abundances of Si and other
lighter elements difficult.

\subsection{Azimuthal variation}

To look into the spectral variation in different azimuthal directions,
the data in the radial range $4' - 60'$ were divided into north, west,
south and east sectors and the resultant pulse-height spectra are
shown in figure 6.  Each sector covers an opening angle of $90^\circ$.
The spectra show that the Fe-K line equivalent widths are different
among the sectors, and the north and the east sectors show relatively
strong iron lines. This feature strongly suggests that the metal
abundance is significantly different along  the azimuthal direction.

Based on the above spectral feature, we examined the radial variation
of the hot-component spectrum for individual sectors.  The ROSAT image
was used again to obtain the ``isothermal'' response, 
and we fit the spectra with single temperature MEKAL
models. Since the statistics of the data were much lower,
2-temperature models resulted in large uncertainties of the spectral
parameters. Therefore, we have used the data only in the energy range above
 2 keV and fit the
data with single temperature models, which minimized the possible
contamination from the cool component.

For each sector, the obtained metal abundance and temperature are
plotted as a function of radius in figure 7, with the parameter values
shown in table 4. For comparison, the parameters for the sum of the 4
sectors are shown in table
4. We note in figure 4 that the single-temperature fit gives the
consistent values to the 2-temperature fit, in the range $r > 20'$ for
the temperature and in all radii for the abundance, respectively. In
the azimuthally divided fit in $r = 36'-44'$, the systematic errors of
the temperature and metal abundance for a change of the background flux 
by 5\% are 0.3 keV and 0.01 solar, respectively.  Among  the temperature
profiles, the west sector shows a peak of about 8 keV in $r = 12'-20'$,
significantly higher than the other 3 data. In the outer regions, the
north and the east sectors show a temperature jump at $r=
36'-44'$. The significance for the temperature jump in the east is
rather small, but at least the north point significantly deviates from
the smooth declining trend of the temperature with radius.  The
detailed spatial variation of the temperature would be seen if a full
2-dimensional map of the ICM temperature is constructed, however this
study requires extensive simulation of the ASCA mapping observation
and is beyond the scope of the present paper.

As for the metal abundance, the sectors except for the east show a
gradual decline with radius, similarly to the sum of all
sectors. 
For each sector,  constant abundance fits give  $\chi^2$ values of 
14.8, 6.95, 10.9, 19.2 with  6 DOF for north, east, south and west 
sectors respectively. Therefore, the abundance variation is significant 
at the 90 \% confidence in 3 sectors excluding the east one.
The east sector shows a significantly high value of
0.61 (0.42-0.82) solar at $r=36'-44'$, which is 2--3 times
higher than the levels in  other 3 sectors. 
The north sector, for example,  shows a very low 
abundance of 0.11 (0-0.26) solar at this radius.  This is the same radius
where the  high temperatures are seen in the north and the east. In other
radii, 90\% errors for  all the sectors  overlap. Therefore, only the
east sector shows an unusual abundance behavior in this limited radial
range. Note that these 4 sectors are symmetrically located around the
cluster center, from which most of the flux contamination
comes. Therefore, systematic differences in the response functions are 
small, and  variation of 
the contaminating flux from the central $4'$ region among the 4
sectors amounts to  only 2\% of the measured flux.
 This assures that the observed difference in the metal
abundance is the real feature. To summarize the azimuthally divided
analysis, both temperature and metal abundance show anomalous 
values in the same radial range ($36'-44'$) in the similar direction
(north and east).

The gain parameters for  the spectral fits  shown in table 4 
tend to be smaller than 
the unity, which was also pointed out by Dupke and Bregman (2001).
However, the values  are consistent to be  the unity within the absolute 
gain accuracy of 1.2\% except for the south sector at  $r=4'-8'$
 (0.977-0.985). 
As  described in Makishima et al.(1996), the GIS gain varies largely  
with temperature by about 1\% K$^{-1}$. Also, the temperature data from the 
satellite are quantized with a step of   0.55 K, which gives a gain error 
$\leq 0.5\%$. 
The long-term gain variation and  the relationship between temperature 
and the gain are continuously monitored and shows a general 
scatter of  0.4\%. 
Furthermore, the systematic uncertainty in  the 
spatial variation of the gain in the GIS field is about 1\% 
(see http://heasarc.gsfc.nasa.gov/docs/asca/cal\_problem.html).  
By adding up these results, the GIS gain accuracy  
of  1.2--1.4\% derived by 
Makishima et al. (1996) is considered as a very realistic estimate.
We also looked into  the GIS intrinsic spectral features at   
Au M-edge and Xe L-edge, 
however, the statistics was not enough to  determine 
whether the Fe line shift was due to the 
gain variation or the redshift of the emission.

\section{Discussion}

The multi-pointing observations of the Perseus cluster from ASCA have
shown significant variations in temperature and metal abundance over a
large scale in the cluster. This is the first detection of such a
large-scale variation in the spectral parameters in this cluster.

\subsection{Spectral Softening in the Central Region}
The energy spectra in the inner regions ($r < 20'$) are significantly
softer than those in outer regions and require two temperature
components. The cool component has a temperature $ kT \sim 2$ keV, and
the hot-component spectrum also shows a gradual softening with a
temperature drop from 8 keV to 5.7 keV toward the center (see table 3
and figure 4). Fabian et al.\ (1994) showed that the SIS spectrum at
the center of the Perseus cluster could be fitted either with a
2-temperature model ($kT =5.62$ keV and 2.05 keV) or with a cooling
flow model, which is consistent with the present result.

The cool emission has a luminosity of $(3.7\pm 0.9) \times 10^{44}$ erg
s$^{-1}$ estimated for an energy range 0.1--10 keV and integrated out 
to $r = 28'$. This is
about 1/4 of the total X-ray luminosity of the cluster. This
cool-to-total luminosity ratio is similar to the value obtained for
the Centaurus cluster (Ikebe et al.\ 1999).  The minimum mass of the
cool component can be estimated by assuming a uniform region with the
highest possible electron density (i.e.\ the central value). Taking
$n_e = 0.015$ after Ettori et al.\ (1998), the minimum mass of the
cool component within a radius of $28'$ becomes $\sim 3 \times
10^{12}M_\odot$. This is much larger than the mass of the interstellar
matter in the central cD galaxy. The observed luminosity and
temperature of the cool component fall on the extension of the $L_{X} -
kT $ relation derived for clusters of galaxies (see e.g.\ Mulchaey
2000), as if the cool component were a relaxed system with its own
gravitational potential.  This is probably not a coincidence since
Ikebe et al.\ (2001) reported that the cool components in several
clusters of galaxies he investigated also showed the consistent
$L_{X}-kT$ relation with general clusters of galaxies.

Another significant feature in the central region is the softening of
the hot-component spectrum toward the center within $r = 8'$. This
trend is contrary to the Virgo cluster case, where the temperature of
the hot component increases at the center (Shibata et al.\ 2000). At
$r=8'$, the electron density is $n_e \sim 3 \times 10^{-3}$ cm$^{-3}$
(Ettori et al.\ 1998) and the cooling time due to the radiation
becomes $t_{\rm cool} \sim 2.4 \times 10^{10}$ yr.  The radiation
cooling could drop the temperature by 20\% in a few Gyr, if there were no heat
input through conduction from the surrounding region. Since the time
scale for the thermal conduction in the central 250 kpc is only
$10^{8}$ years (Spitzer 1956), suppression of the conductivity to 
less than 1\% of the standard value would be necessary 
to maintain the temperature
gradient (see also Ettori and Fabian 2000).

Recent XMM-Newton study of A1795 reveals that the temperature within
$10'$ ($\sim 1$ Mpc) from the center gradually drops from $kT=6.4$ keV
to $2.2-2.4 $ keV toward the center (Tamura et al.\ 2001). For this
cluster, two-temperature fit for the ASCA spectrum within $3'$ from
the center showed $kT = 6.54^{+0.56}_{-0.50}$ keV and
$1.70^{+0.24}_{-0.26}$ keV (Xu et al. 1998), which approximately
correspond to the maximum and the minimum temperatures in the observed
field. Therefore, because of the insufficient angular resolution of
ASCA XRT, the two-temperature feature in the GIS spectrum of the
Perseus cluster is likely to indicate the range of temperature within
the observed field which contains a strong temperature gradient.

\subsection{Temperature Structure in the Outer region} 
In the outer region, the radial temperature profile in figure 4 shows
a peak ($kT \sim 8.5$ keV) at around $15'$ and then drops to $\sim 7$
keV in the outermost region. Markevitch et al.\ (1998) reported a
universal temperature decrease with radius for 30 clusters,
characterized by a factor of 2 drop at half the virial radius ($r_{\rm
vir}$). For the Perseus cluster, $0.5 r_{\rm vir}$ corresponds to 1.7
Mpc ($r=55'$) and the observed temperature profile does not agree with
a drop as much as factor of 2 at this radius. Recent results with ASCA
(Kikuchi et al.\ 1999) and BeppoSAX (Irwin \& Bregman 2000) for other
clusters also indicate that the radial temperature profiles are more
closer to being isothermal.

The non-radial temperature structures in the outer regions have been
studied from Spartan (Snyder et al.\ 1990), ASCA (Arnaud et al.\
1994), and ROSAT (Ettori et al.\ 1998), which all reported significant
temperature variation.  In particular, ROSAT PSPC observation
indicated a cool emission with $kT \sim 3$ keV in the east between
$20'$ and $50'$ from the center (Ettori et al.\ 1998). This region
also showed an enhanced X-ray surface brightness with significant
deviation from a smooth elliptical distribution, which suggested a
subcluster merger (Schwarz et al.\ 1992). The ASCA temperatures in the
four sectors indicate a fairly large scatter in the range $12' - 20'$
and $36'-44'$ (see figure 7). In the inner region ($12'-20'$), the
north sector is significantly hotter than other sectors. This may
correspond to the hot region reported by Arnaud et al.\ (1994) based
on the GIS data. However, these early ASCA results were not fully corrected
for the energy dependent response of the X-ray telescope and may
contain significant systematic errors. We also note that the reported
cool region at $20'-50'$ in the east (Ettori et al.\ 1998) is not
clearly recognized in the GIS data. Suppression of the data in the
energy range below 2 keV in the analysis may have deprived our
sensitivity to the cool emission. It is suggestive that, within
$20'-28'$, the east sector indicates the lowest temperature.

The north and west sectors exhibit some hotter emission around $40'$,
which is close to the outer boundary of the residual emission when the
smooth $\beta$-model is subtracted from the surface brightness data
(Ettori et al.\ 1998).  The metal abundance is high in the east and
low in the north.  The high metallicity suggests an enhanced star
formation activity in the galaxies triggered by merger shocks, however
with the present coarse resolution analysis the detailed feature seems
to remain unresolved.

\subsection{Metal Distribution} 
For the metal abundance, the GIS shows the central abundance within
$4'$ to be $0.44 \pm 0.03$ solar.  Our value is lower than the results
from Spacelab 2 (0.7 solar for no point-source case, Ponman et al.\
1990) and Spartan 1 (0.8 solar, Ulmer et al.\ 1987; Kowalski et al.\
1993), but consistent with the Einstein SSS (0.44 solar, Mushotzky et
al.\ 1981) and BBXRT ($\sim 0.5$ solar, Arnaud et al.\ 1992)
results. Recent BeppoSAX observation also indicated 0.48 solar in the
central $2'$ region (Molendi et al.\ 1998). The present result is the
first to show the metal abundance out to $1^\circ$ with enough
sensitivity. The abundance profile shows a gradual decline except for
a jump at about $40'$ in the east and the outermost level is $0.18 \pm
0.08$ solar. De Grandi and Molendi (2000) showed that the profile
almost traces the optical light distribution of early type galaxies,
but was slightly broader than the predicted ones. If the metals were
to drift by about 50 kpc the profile can be explained. 

It has been pointed out that the hot gas in the cluster center is
opaque to a resonant scattering of Fe K$\alpha$ line (Gilfanov,
Sunyaev, Churazov 1987), if there is no bulk motion of the gas. The
scattering tends to smooth out the radial profile of the line
intensity by conserving the total number of line photons. This effect
was studied for the Perseus cluster with ASCA by Akimoto et al.\
(1999) and with BeppoSAX by Molendi et al.\ (1998). They predict that
within $4'$ from the center, the abundance estimated from Fe K$\alpha$
line is lower by a factor of about 2. If this is the case, as
suggested from the Fe K$\beta$ line profile observed with BeppoSAX
(Molendi et al.\ 1998), the central abundance should be about 0.9
solar which results in a strong gradient in the metal abundance.  The
dashed line in the bottom panel of figure 4 shows the expected
abundance profile when the metal mass density follows a $\beta$ model
with $\beta = 1.0$ and core radius of $11'$, respectively, which
represents the distribution of cluster galaxies (Kent, Sargent 1983;
Eyles et al.\ 1991). Therefore, in the Perseus cluster it is likely
that the metal mass density traces the galaxy distribution, consistent
with the feature previously observed in AWM~7 (Ezawa et al.\ 1997). To
confirm the significance of the resonance scattering effect,
observation of the cluster center with better energy resolution and
angular resolution would be required (see also B\"ohringer et al.\
2001).

Finally, the GIS  spectra  suggest a Fe-K line energy shift by about 1\% 
in the south sector, similar to the feature  reported by 
Dupke and Bregman (2001). However, the error range of the 
energy shift is smaller than the the absolute gain accuracy (1.2\%) 
of the GIS, and we conclude there is no significant velocity gradient in the 
Perseus cluster.

\bigskip
The authors thank  Dr.\ H.~B\"ohringer and Dr.\ Y. Ikebe for
useful discussion and the support in analyzing the ROSAT data.
This work was partly supported by the Grants-in Aid for
Scientific Research No.\ 08404010 and No.\ 12304009 from the Japan
Society for the Promotion of Science.

\section*{References}

\re
 Akimoto F., Furuzawa A., Tawara Y., Yamashita K. 1999, AN 320, 283
\re
Allen S. W., Edge A. C., Fabian A. C., B\"ohringer H., Crawford
 C. S., Ebeling H., Johnstone R. M., Naylor T., Schwarz
 R. A. 1992, MNRAS 259, 67
\re
Anders E., G`revesse N. 1989, Geochemica et Cosmochemica Acta 53, 197
\re
Arnaud K. A., Mushotzky R. F., Serlemitsos P. J., Boldt E., Holt
 S. S., Jahoda K., Marshall F. E., Petre R. et al.\ 1992 in Proc.\
 Ginga Memorial Symp., ed.\ F. Makino \& F. Nagase (Tokyo: ISAS) 114
\re
Arnaud K. A., Mushotzky R. F., Ezawa H., Fukazawa Y., Ohashi T.,
 Bautz M. W., Crewe G. B., Gendreau K. C. et al.\ 1994, ApJ 436, L67
\re
W., Fabian A.C., Edge A.C., Neumann D.M. 1993, MNRAS, 264, L25
\re B\"ohringer, H., Belsole, E., Kennea, J., Matsushita, K.,
Molendi, S., Worrall, D. M., Mushotzky, R. F., Ehle, M., et al.\ 2001,
A\&A, 356, L181
\re
Churazov E., Gilfanov M., Forman W., Jones C. 1999, ApJ 520, 105
\re
Churazov E.,Forman W., Jones C., B\"ohringer H. 2000, A\&A 356, 788 
\re
De Grandi S., Molendi S. 2000, ApJ accepted astro-ph/0012232
\re
Dupke R., Arnaud K. 1999, AN 320, 284
\re
Dupke R.A., Bregman J.N. 2001, ApJ 547, 705 
\re
Ettori S., Fabian A. C., White D. A. 1998, MNRAS 300, 837
\re
\re
Eyles C. J., Watt M. P., Bertram D., Church M. J., Ponman T. J.,
 Skinner G. K., Willmore A. P. 1991, ApJ 376, 23
\re
Ezawa H., Fukazawa Y., Makishima K., Ohashi T., Takahara F., Xu
 H., Yamasaki N. Y. 1997, ApJ 490, L33
\re
Fabian A. C., Arnaud K. A., Bautz M. W., Tawara Y. 1994, ApJ 436,
L63
\re
Fabian A.C., Sanders J.S., Ettori S., Taylor G.B., Allen S.W., 
 Crawford C.S., Iwasawa K., Johnstone R.M., Ogle P.M. 2000, MNRAS 318, L65
\re
Fukazawa Y., Ohashi T., Fabian A. C., Canizares C. R., Ikebe Y.,
 Makishima K., Mushotzky R. F., Yamashita K. 1994, PASJ 44, L55
\re
Furusho T., Yamasaki N.Y., Ohashi T., Shibata R., Kagei T., 
 Ishisaki Y.,  Kikuchi K., Ezawa H., Ikebe Y. 2000, PASJ accepted
\re
Gilfanov M. R., Sunyaev R. A., Churazov E. M. 1987, Soviet
Astronomy Letters 13, 3
\re
Honda H., Hirayama M., Watanabe M., Kunieda H., Tawara Y.,
 Yamashita K., Ohashi T., Hughes J. P., Henry J. P. 1996, ApJ 473, L71
\re
Ikebe Y., Ph.D thesis, 1995, University of Tokyo 
\re
Ikebe Y., Makishima K., Ezawa H., Fukazawa Y., Hirayama M., Honda
 H., Ishisaki Y., Kikuchi K. et al.\ 1997, ApJ 481, 660
\re
Ikebe Y., Makishima K., Fukazawa Y., Tamura T., Xu H., Ohashi T.,
 Matsushita K. 1999, ApJ 525, 58
\re
 B\"ohringer, H., \& Tanaka, Y., presentaion at ``New Century of X-ray
 Astronomy'', Yokohama, March, 2001\\ or private communication ?
\re
Ishisaki Y., Ueda Y., Kubo H., Ikebe Y., Makishima K. and the GIS team
 1997 ASCA News No 5, 26
\re
Irwin J.A., Bregman J.N. 2000, ApJ 538, 543  
\re
Kent S. M., Sargent W. L. W. 1983, AJ 88, 697
\re
Kikuchi K., Furusho T., Ezawa H., Yamasaki N. Y., Ohashi T.,
  Fukazawa Y., Ikebe Y. 1999, PASJ 51, 301
\re
Kowalski M. P., Cruddace R. G., Snyder W. A., Frits G. G., Ulmer
 M. P., Fenimore E. E. 1993, ApJ 412, 489
\re
Lampton M., Margon B., Bowyer S. 1976, ApJ Letter 208, 177  
\re
Makishima K., Tashiro M., Ebisawa K., Ezawa H., Fukazawa Y., Gunji
 S., Hirayama M., Idesawa E. et al.\ 1996, PASJ 48, 171
\re
Markevitch M., Forman W.  R.; Sarazin C. L., Vikhlinin A. 1998,
 ApJ 503, 77
\re
Molendi S., Matt G., Antonelli L. A., Fiore F., Fusco-Femiano R.,
 Kaastra J., Maccarone C., Perola C. 1998, ApJ 499, 608
\re
Mushotzky R. F., Holt S. S., Smith B. W., Boldt E. A., Serlemitsos
 P. J. 1981, ApJ 244, L47
\re
Ohashi T., Ebisawa K., Fukazawa Y., Hiyoshi K., Horii M., Ikebe
 Y., Ikeda H., Inoue H.  Ishida M. 1996, PASJ 48, 157
\re
Peterson J.R., Paerels F.B.S., Kaastra J.S., Arnaud M., Reiprich T.H., 
Fabian A.C., Mushotzky R.F., Jernigan J.G., Sakelliou I. 2000, A\&A Letters 
365, 104
\re
Ponman T. J., Bertram D., Church M. J., Eyles C. J., Watt M. P.,
 Skinner G. K., Willmore A. P. 1990, Nature 347, 450
\re
Mulchaey J.S. 2000\ Ann. Rev. Astron. Astrophys., 38, 289
\re
Schwarz R. A., Edge A. C., Voges W., B\"ohringer H., Ebeling H.,
 Briel U. G. 1992, A\&A  256, L11
\re
Shibata R., Matsusita K., Yamasaki N.Y., Ohashi T., Ishida M., 
 Kikuchi K., B\"ohringer H., Matsumoto H. 2001, ApJ 549, 228
\re
Snyder W. A., Kowalski M. P., Cruddace R. G., Fritz G. G.,
 Middleditch J., Fenimore E. E., Ulmer M. P., Majewski S. R. 1990, ApJ
 365, 460
\re Spitzer, L. 1956, Physics of Fully Ionized Gases,
Interscience Publishers, New York
\re 
Takahashi T., Markevitch M., Fukazawa Y., Ikebe Y., Ishisaki Y.,
 Kikuchi K., Makishima K., Tawara Y. et al.\ 1995, ASCA News 3, 24
\re
Tamura T., Kaastra J.S., Peterson J.R., Paerels F., Mittaz P.D., 
Trudolyubiv S.P., Stewart G., Fabian A.C. et al.\ 2001 A\&A Letters  365, 87
\re
Tanaka Y., Inoue H., Holt S. S. 1994, PASJ 46, L37
\re
Ulmer M. P., Cruddace R. G., Fritz G. G., Snyder W. A., Fenimore
 E. E. 1987, ApJ 319, 118
\re
Watanabe M., Yamashita K., Furuzawa A., Kunieda H., Tawara, Y.,
 Honda, H. 1999, ApJ 527, 80
\re
Xu H., Makishima K., Fukazawa Y., Ikebe Y., Kikuchi K., 
Ohashi T., Tamura T. 1998, ApJ  500, 538

\begin{figure}
\centerline{\psfig{file=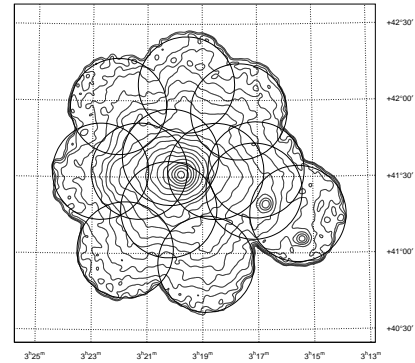,width=60mm}}
\caption{The X-ray image of the Perseus cluster observed with the ASCA
GIS. Each pointed field is shown with a circle of $40'$ diameter,
representing the GIS field of view. The energy range is 0.7--10 keV.}
\label{fig:image}
\end{figure}

\begin{figure}
\centerline{\psfig{file=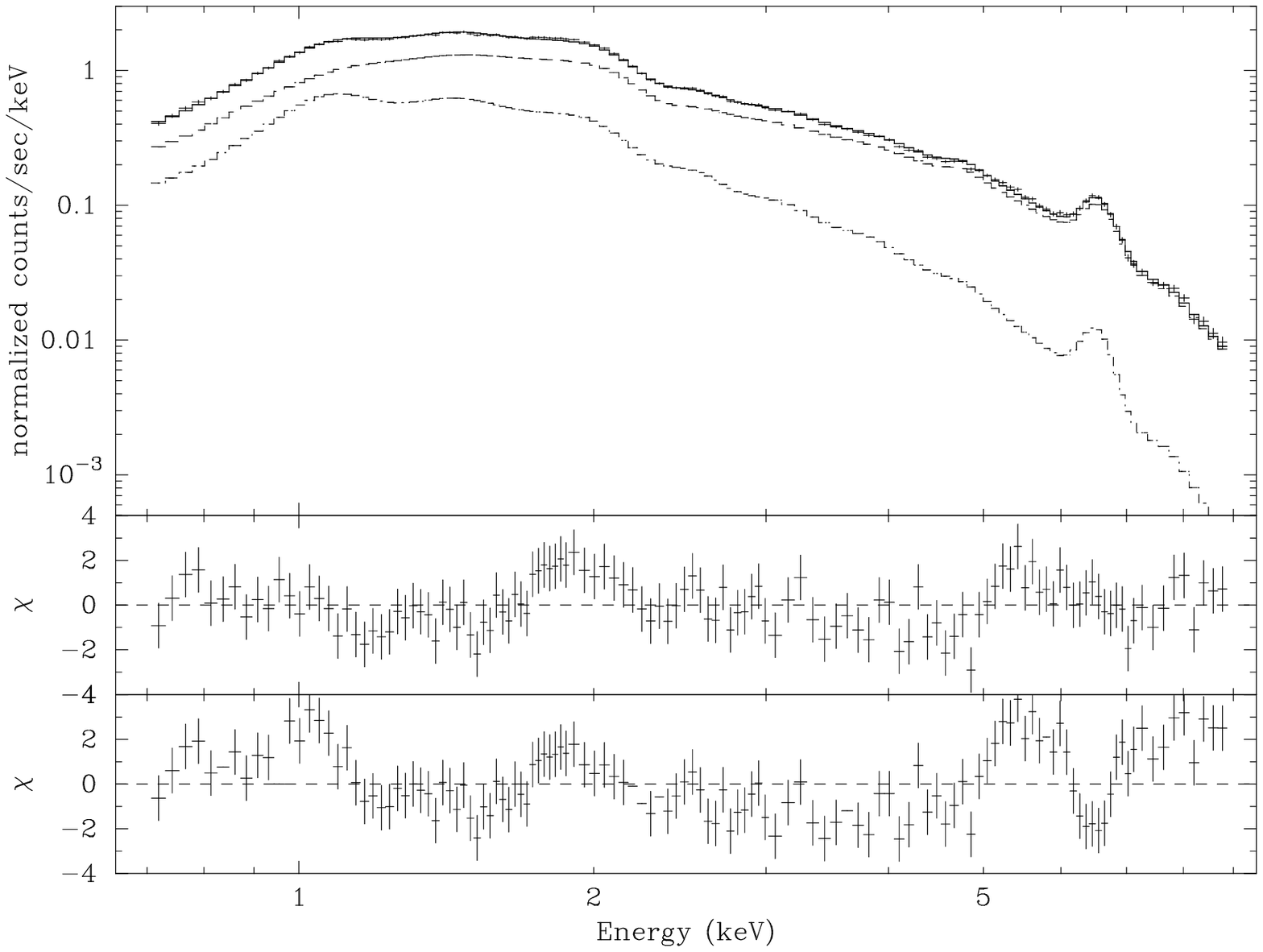,width=60mm}}
\vspace{2mm}
\centerline{(a)}

\centerline{\psfig{file=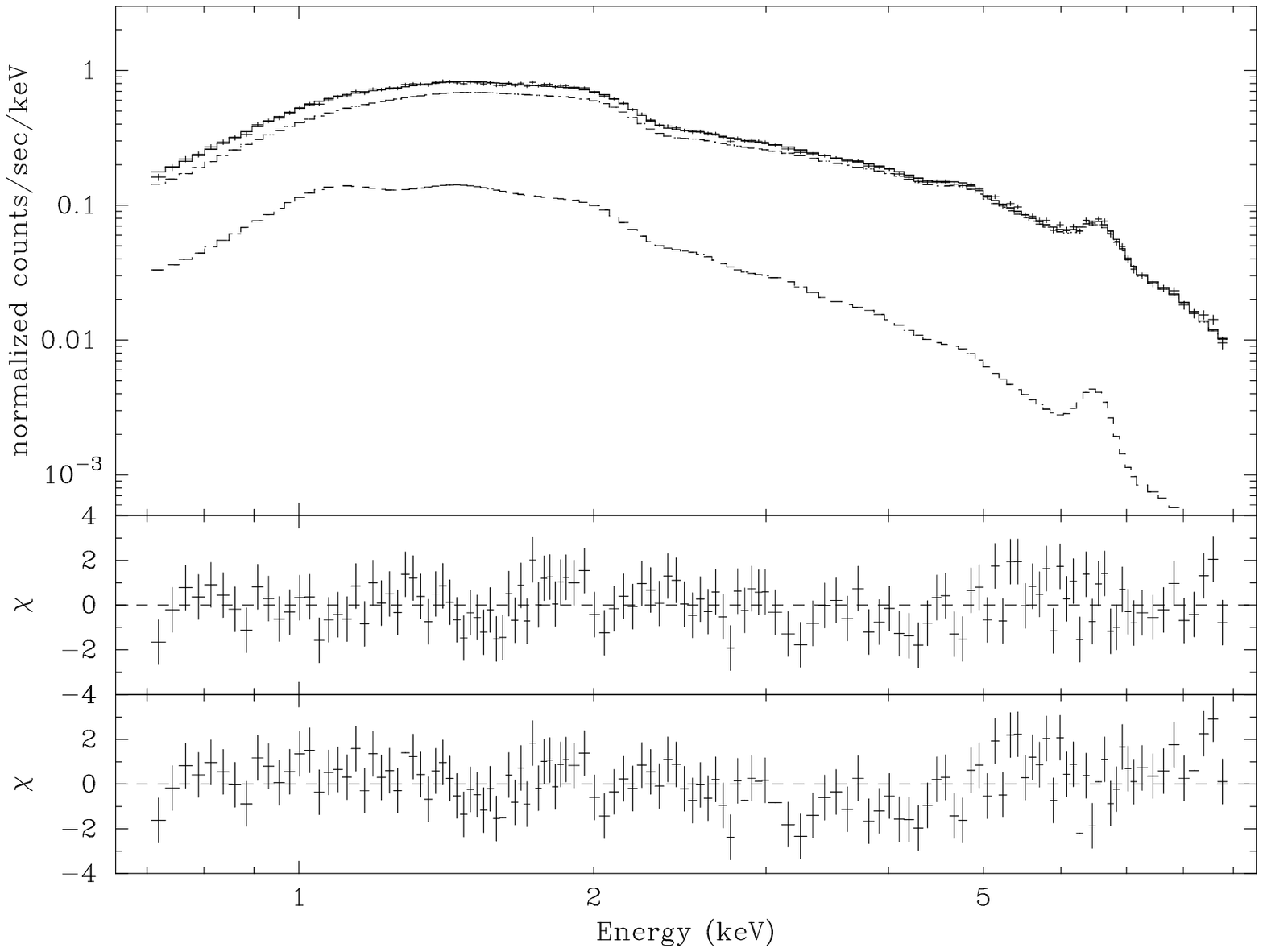,width=60mm}}
\vspace{2mm}
\centerline{(b)}

\caption{Pulse-height spectra  for the central region ($r < 4'$) (a) 
and the annular region ($12'<r<20'$) (b) of the
Perseus cluster for the sum of the GIS2 and GIS3 data. The
2-temperature model with fixed $N_{\rm H}$ and common abundance (the
first row in table 2) are  fitted to the data, and the middle  panels
shows the residual of the fit. The bottom panels shows the residuals 
when the data are fitted by 1-temperature model with the same $N_{\rm H}$.}
\label{fig:spectralfit}
\end{figure}

\begin{figure}
\centerline{\psfig{file=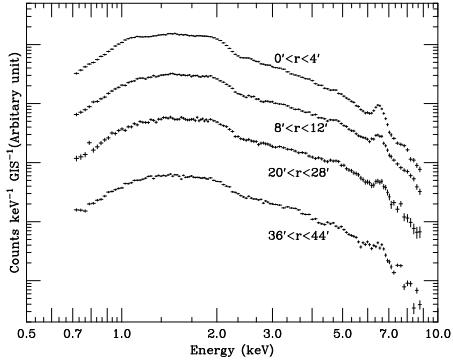,width=60mm}}
\caption{Energy spectra obtained by the GIS for 4 concentric annular regions
in the Perseus cluster. The gradual decrease of the Fe K-line
equivalent width as a function of radius is suggested.}
\label{fig:ring-spectra}
\end{figure}

\begin{figure}
\centerline{\psfig{file=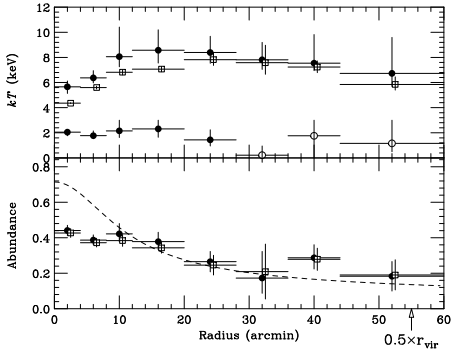,width=60mm}}
\caption{Radial variation of the temperature (top) and metal abundance
(bottom) obtained from the annular spectral fits. The filled circles
indicate results with the two-temperature model, the open circles 
are best-fit temperatures of the cool component whose significance 
are less than 99\%.  The open squares
are for single-temperature fits in the energy range above 2 keV\@.
The dashed line in the bottom panel shows an abundance profile when
the metal mass density follows a $\beta$ model with $\beta = 1.0$ and
core radius of $11'$, respectively. }
\label{fig:radial}
\end{figure}

\begin{figure}
\centerline{\psfig{file=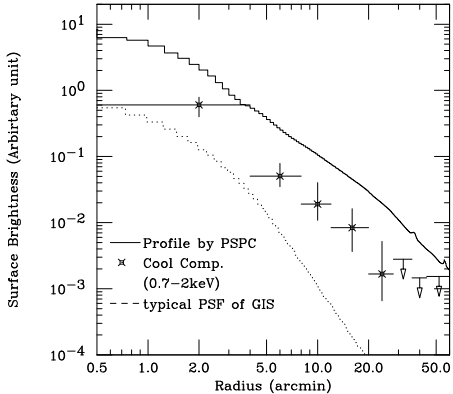,width=60mm}}
\caption{Surface brightness distribution of the cool component, compared with
the PSPC radial profile and the point spread function of the GIS
instrument. The emission is extended.}
\label{fig:cool}
\end{figure}

\begin{figure}
\centerline{\psfig{file=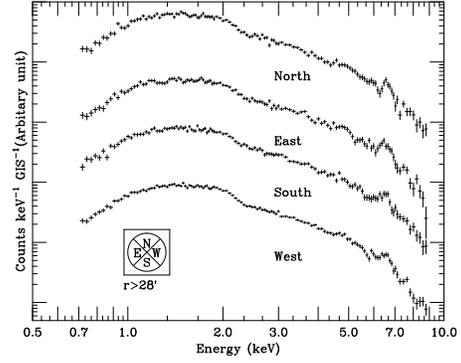,width=60mm}}
\caption{Pulse-height spectra for the four $90^\circ$ sectors in the radius
range $28'-60'$. The Fe line equivalent widths are different among the
sectors.}
\label{fig:sector-spectra}
\end{figure}

\begin{figure}
\centerline{\psfig{file=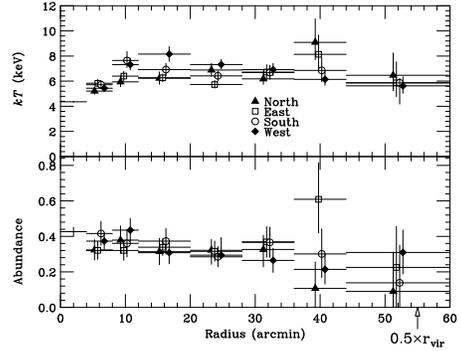,width=60mm}}
\caption{Radial variation of temperature and metal abundance in the 4 sectors.}
\label{fig:sector}
\end{figure}

\begin{table}
\caption{Summary of ASCA observations}
\label{tab:obslog}
\begin{center}

\begin{tabular}[h]{l|l|l|l|l}
Sequential  & Observation & Exposure &\multicolumn{2}{l}{FOV center}     \\ 
No.    & Start (UT)  & (sec.) & $\alpha_{2000}$ & $\delta_{2000}$     \\
\hline  \hline
80007000   &93/08/06 04:19 & 11737& 50.040 & 41.619  \\ 
80008000   &93/08/06 18:30 & 18449& 49.630 & 41.537  \\ 
80009000   &93/09/15 00:39 & 16978& 49.264 & 41.546  \\ 
81026000   &94/02/09 17:13 & 19862& 49.284 & 41.407  \\ 
81027000   &94/02/10 03:10 & 16935& 48.888 & 41.258  \\ 
83051000   &95/08/18 20:06 & 10307& 50.325 & 41.538  \\ 
83052000   &95/08/19 12:11 & 15817& 50.677 & 41.537  \\ 
83053000   &95/09/04 13:06 & 13330& 50.057 & 41.287  \\ 
85000000   &97/02/14 11:22 & 18233& 50.516 & 41.961  \\ 
85001000   &97/02/14 23:51 & 18108& 49.895 & 42.103  \\ 
85002000   &97/02/15 12:42 & 15501& 49.423 & 41.923  \\ 
85003000   &97/02/16 19:50 & 17974& 50.449 & 41.023  \\ 
85004000   &97/02/17 09:11 & 18490& 49.721 & 40.930  \\ 
\end{tabular}
\end{center}
\end{table}

\clearpage
\begin{table}
\caption{Spectral fits for the central region within the radius of $4'$.}
\label{tab:fit-center}
\begin{center}

\begin{tabular}[h]{l|l|l|l|l|l|l}
$N_{\rm H hot}$ & $kT_{\rm hot}$ & $Z_{\rm hot}$ & $N_{\rm H cool}$ & $kT_{\rm cool}$ & $Z_{\rm cool}$ &$\chi ^{2}$/dof \\ 
(10$^{21}$cm$^{-2}$) & (keV)   &          &  (10$^{21}$cm$^{-2}$) & (keV)   &          &              \\
\hline \hline
1.4(fix) &5.73(5.15-6.23)&0.442(0.413-0.470)&$=N_{\rm H hot}$ &2.05(1.76-2.26)&$=Z_{\rm hot}$ & 149.8/121 \\ 
1.4(fix) &5.66(5.21-6.22)&0.458(0.422-0.503)&$=N_{\rm H hot}$ &1.91(1.71-2.21)&0.342(0.250-0.664)& 148.0/120 \\
1.63(1.42-1.84)&5.43(5.08-6.04)&0.432(0.422-0.503)&$=N_{\rm H hot}$ &1.75(1.56-2.14)&$=Z_{\rm hot}$ & 146.7/120 \\  
2.03(1.62-2.32)&5.32(5.06-5.57)&0.449(0.422-0.477)&$=N_{\rm H hot}$ &1.37(1.27-1.73)&0.223(0.170-0.291) &138.8/119 \\
1.4(fix) &5.27(5.04-5.52)&0.440(0.416-0.465)&4.92(3.58-6.28)  &1.30(1.17-1.40)&$=Z_{\rm hot}$ &132.7/120 \\
1.4(fix) &5.23(4.97-5.50)&0.436(0.409-0.463)&4.92(3.58-6.28)&1.29(1.16-1.40)&0.548(0.328-1.14)&132.3/119\\
\end{tabular}
\vspace{6pt}

\noindent
Note: $Z$ represents metal abundance, and the values in parenthesis indicate single-parameter 90\% limits.
\end{center}
\end{table}

\begin{table}
\caption{Spectral fits for the annular regions}
\label{tab:fit-cnnular}
\begin{center}

\begin{tabular}[h]{l|l|l|l|l|l|l}
radius& $kT_{\rm hot}$ (keV) & $Z$  & $kT_{\rm cool}$ (keV)  & Cool fraction (\%)& gain &$\chi ^{2}$/dof \\ 
\hline \hline
0-4  & 5.73(5.15-6.23)& 0.442(0.413-0.470)& 2.05(1.76-2.26)& 34.2(22.6-44.7) &0.996(0.993-0.998)  &149.96/121 \\
     & 4.09(4.05-4.12)& 0.496(0.474-0.519)& --             & --              &0.994(0.993-0.994)& 304.52/123\\
4-8  & 6.37(5.97-6.95)& 0.387(0.361-0.416)& 1.77(1.53-2.15)& 18.5(12.8-28.8) &0.999(0.995-1.0)& 137.08/121 \\
     & 5.21(5.15-5.28)& 0.411(0.388-0.434)& --             & --              &0.994(0.993-0.995) &235.78/123\\
8-12 & 8.06(7.25-10.4)& 0.422(0.377-0.480)& 2.16(1.63-3.0) & 17.8(9.2-38.6)  &0.993(0.990-0.993) &111.54/121 \\
     & 6.42(6.30-6.55)& 0.411(0.380-0.442)& --             & --              &0.990(0.988-0.992) &152.35/123\\
12-20& 8.57(7.55-10.2)& 0.378(0.333-0.431)& 2.31(1.64-3.0) & 18.2(7.9-35.3)  &0.993(0.990-0.994) &117.29/121 \\
     & 6.79(6.66-6.91)& 0.357(0.328-0.386)& --             & --              &0.990(0.989-0.992) &149.46/123 \\
20-28& 8.39(7.53-9.68)& 0.265(0.208-0.323)& 1.44(0.95-2.24)& 8.5(3.4-26.4)   &0.993(0.989-1.0)   &111.98/121  \\
     & 7.34(7.11-7.60)& 0.277(0.225-0.330)& --             & --              &0.989(0.989-0.994) &123.93/123 \\
28-36& 7.81(7.05-9.20)& 0.173(0.087-0.323)& 0.22(0.09-0.95)& 6.7(0.0-30.2)   &0.998(0.978-1.01)  &85.97/121 \\
     & 7.94(7.20-8.86)& 0.208(0.054-0.364)& --             & --              &0.986(0.973-0.997) & 87.87/123\\
36-44& 7.54(6.95-9.82)& 0.287(0.224-0.361)& 1.76(0-3.0)    & 4.0(0.0-27.7)   &0.983(0.978-0.991)&108.00/121 \\
     & 7.14(6.88-7.40)& 0.291(0.231-0.352)& --             & --              &0.983(0.979-0.987)&109.12/123\\
44-60& 6.73(5.70-9.59)& 0.183(0.100-0.266)& 1.16(0.50-3.0) & 5.6(0.0-56.0)   &0.988(0.980-1.0)   &76.50/121 \\
     & 5.89(5.64-6.24)&0.200(0.122-0.280) & --             & --              &0.983(0.980-0.991) &79.39/123\\
\end{tabular}
\vspace{6pt}

\noindent
Note: 0.7-9 keV, $N_{\rm H}$ is fixed to the Galactic value $1.4\times
10^{21}$ cm$^{-2}$. One and two-component MEKAL model is used. Cool
fraction is the ratio of the cool-component to the total fluxes in the
energy range 0.7 -- 2 keV.
\end{center}
\end{table}
 
\clearpage
\begin{table}
\caption{Spectral fits for the four sectors }
\label{tab:fit-sector}
\begin{center}

\begin{tabular}[h]{l|l|l|l|l}
radius(arcmin)& $kT$ (keV) & $Z$ & gain &  $\chi ^{2}$/dof \\ 
\hline \hline
\multicolumn{3}{l}{All sectors} \\ \hline
0-4   & 4.36(4.29-4.43) &0.427(0.402-0.451)&0.9996(0.9988-0.9997) &129.68/70 \\
4-8   & 5.60(5.49-5.72) &0.372(0.348-0.397)&0.999(0.998-1.0) & 104.88/70 \\
8-12  & 6.82(6.61-7.03) &0.384(0.352-0.417)&0.998(0.995-1.0) & 76.72/70 \\
12-20 & 7.07(6.86-7.28) &0.343(0.313-0.374)&0.998(0.993-0.999) & 82.62/70 \\
20-28 & 7.82(7.36-8.36) &0.246(0.190-0.301)&0.999(0.990-1.004) & 53.81/70    \\     
28-36 & 7.58(6.56-9.07) &0.209(0.055-0.364)&0.986(0.967-1.011) & 43.40/70 \\
36-44 & 7.23(6.79-7.72) &0.279(0.216-0.341)&0.988(0.978-0.994) & 79.26/70 \\
44-60 & 5.85(5.40-6.45) &0.190(0.106-0.276)&0.989(0.976-1.0)   & 43.82/70  \\
\hline
\multicolumn{3}{l}{North} \\ \hline
4-8   &5.21 (4.95-5.48) &0.323(0.267-0.381)& 1.000(0.995-1.007) &69.70/70\\
8-12  &5.94 (5.57-6.44) &0.379(0.304-0.459)& 0.999(0.991-1.002)& 81.35/70\\
12-20 &6.23 (5.76-6.74) &0.314(0.240-0.389)& 0.990(0.989-1.001)& 67.98/70\\ 
20-28 &6.89 (6.41-7.39) &0.322(0.243-0.383)& 1.000(0.990-1.008) & 72.11/70\\
28-36 &6.18 (5.75-6.72) &0.326(0.228-0.406)& 1.001(0.990-1.007) & 82.85/70\\
36-44 &9.09 (7.50-10.96)&0.107(0.0  -0.256)& 0.989(0.964-1.007) & 71.83/70\\
44-60 &6.47 (5.25-8.24) &0.090(0.0  -0.311)& 1.001(0.965-1.029) & 98.48/70\\
\hline
\multicolumn{3}{l}{East} \\ \hline
4-8   &5.83(5.58-6.13) &0.321(0.271-0.371) &0.998(0.991-1.000) &73.17/70\\
8-12  &6.40(6.04-6.78) &0.321(0.265-0.377) &1.009(1.000-1.011) &69.46/70\\
12-20 &6.27(5.96-6.58) &0.339(0.293-0.386) &0.999(0.994-1.004) &77.40/70\\
20-28 &5.73(5.45-6.05) &0.316(0.260-0.374) &0.996(0.989-1.001) &61.78/70\\
28-36 &6.72(6.16-7.31) &0.369(0.283-0.455) &0.989(0.980-0.998) &68.61/70\\
36-44 &8.14(6.96-9.67) &0.609(0.419-0.815) &0.984(0.972-1.000) &63.73/70\\
44-60 &5.87(4.78-7.54) &0.225(0.0-0.456)   &0.941(0.930-1.031) &57.02/70\\
\hline
\multicolumn{3}{l}{South} \\  \hline
 4-8  &5.72(5.42-6.04) &0.416(0.350-0.484)& 1.000(0.998-1.005) &84.04/70\\
 8-12 &7.65(7.10-8.35) &0.361(0.285-0.437)& 0.985(0.977-0.988) &69.40/70\\
12-20 &6.93(6.45-7.42) &0.373(0.300-0.445)& 1.000(0.990-1.005) &71.91/70\\
20-28 &6.43(6.04-6.83) &0.285(0.229-0.342)& 0.998(0.989-1.003) &83.05/70\\
28-36 &6.69(6.15-7.26) &0.367(0.282-0.452)& 0.997(0.984-1.000) &55.50/70\\
36-44 &6.86(5.97-7.91) &0.301(0.168-0.437)& 0.967(0.955-0.989) &72.30/70\\
44-60 &5.05(4.18-6.21) &0.138(0.0  -0.350)& 0.970(0.929-1.000) &67.03/70\\
\hline
\multicolumn{3}{l}{West} \\ \hline
 4-8  &5.44(5.24-5.65) &0.373(0.328-0.419) &1.000(0.995-1.000) &84.69/70 \\
 8-12 &7.32(6.91-7.83) &0.435(0.371-0.501) &0.988(0.985-1.000) &89.10/70 \\ 
12-20 &8.15(7.58-8.75) &0.308(0.246-0.371) &0.999(0.996-1.002) &91.14/70 \\ 
20-28 &7.33(6.99-7.73) &0.294(0.245-0.344) &0.999(0.995-1.007) &76.85/70 \\ 
28-36 &6.92(6.44-7.41) &0.264(0.197-0.333) &0.993(0.989-1.004) &57.96/70 \\ 
36-44 &6.15(5.67-6.76) &0.214(0.131-0.298) &0.997(0.982-1.007) &56.78/70 \\ 
44-60 &5.62(5.04-6.35) &0.310(0.189-0.436) &0.986(0.972-0.999) &49.79/70 \\ 
\end{tabular}
\vspace{6pt}

\noindent
Note: 2-9keV 1
component,$N_{\rm H}$ is fixed to the Galactic value $1.4 \times
10^{21}$ cm$^{-2}$. 
\end{center}
\end{table}

\end{document}